\documentclass[preprint2]{aastex}
\usepackage{txfonts}
\usepackage{graphicx}

\shorttitle{absolute timing}
\shortauthors{Molkov et al.}

\begin{document}

\title{Absolute timing of the Crab pulsar with the INTEGRAL/SPI telescope}

\author{S. Molkov, E. Jourdain and J.P. Roques}

\email{ molkov@iki.rssi.ru}

\affil{CESR -- Universite de Toulouse (UPS), CNRS (UMR 5187),
9 Av. du Colonel Roche, 31028 Toulouse Cedex 4, France}

\altaffiltext{1}{Based on observations with INTEGRAL,
    an ESA project with instruments
    and science data centre funded by ESA member states (especially the PI
    countries: Denmark, France, Germany, Italy, Switzerland, Spain), Czech
    Republic and Poland, and with the participation of Russia and the USA.}

\begin{abstract}

We have investigated the pulse shape evolution of the Crab pulsar emission in
the hard X-ray domain of the electromagnetic spectrum. In particular, we
have studied the alignment of the Crab pulsar phase profiles
measured in the hard $X-$rays and in other wavebands. To obtain
the hard $X-$ray pulse profiles, we have used six year ($2003-2009$, with a total 
exposure of about 4 Ms) of publicly available data
of the SPI telescope on-board of the INTEGRAL observatory, folded 
with the pulsar time solution derived from the Jodrell Bank Crab Pulsar Monthly
Ephemeris (http://www.jb.man.ac.uk). We found that the main pulse in the hard $X-$ray
$20-100$ keV energy band
is leading the radio one by $8.18\pm0.46$ milliperiods in phase, or
$275\pm15~\mu s$ in time. Quoted errors represent only statistical uncertainties.
Our systematic error is estimated to be $\sim 40 \mu s$ and is mainly caused by
the radio measurement uncertainties.
In hard $X-$rays, the average distance between the main pulse and
interpulse on the phase
plane is $0.3989\pm0.0009$. To compare
our findings in hard $X-$rays with the soft $2-20$ keV X-ray band, we have used
data of quasi-simultaneous Crab observations with the PCA monitor on-board
the Rossi X-Ray Timing Explorer (RXTE) mission. The time
lag and the pulses separation values measured in the $3-20$ keV band are $0.00933\pm0.00016$
(corresponding to $310\pm6~\mu s$) and $0.40016\pm0.00028$ parts of the cycle, respectively.
While the pulse separation values measured in soft $X-rays$ and hard $X-rays$ agree,
the time lags are statistically different.
Additional analysis show that the delay between the radio and X-ray signals varies with energy
in the 2 - 300 keV energy range.
We explain such a behaviour as due to the superposition of two independent components responsible
for the Crab pulsed emission in this energy band.

\end{abstract}

\maketitle

%
%________________________________________________________________

\section{Introduction}

The Crab pulsar (PSR B0531+21) is the best studied  isolated
pulsar. 
%It is located within the Crab Nebula (SN1054, M1) at
%the coordinates $RA=83.633217$ and $DEC=22.014464$ in the J2000 epoch.
The pulsed emission  was discovered long ago and in the X-rays
\citep{fritz69,bradt69} and in the $\gamma$-rays \citep{kurfess71} domains.
and its pulse morphology
has been studied in the full range of the electromagnetic spectrum.
In all energies, the pulse profile has two prominent
features, the main pulse (or the first peak, P1) and the interpulse (or the second
peak, P2). The relative intensities of these peaks depend on the energy band.
The second peak dominates in the $\sim 200 - 1200$ keV energy band.
By all appearances, the distance between the peaks on the
phase plane is almost  constant in time, slightly varying around the value
$\Delta \psi =0.40(0)$ depending on energy  
\citep[see][]{thompson77, wills82,
white85, nolan93, masnou94, moffett96, pravdo97, kuiper01, brandt2003, rots2004,
magic2008}. For a long time it has been assumed that
both peaks are perfectly lined up in phase over the whole energy range. This
assumption has been disputed for the first time in the work of Masnou et al. (1994).
Based on the data of the FIGARO II
telescope (balloon experiment), the authors found that the first peak in the $0.15-4$ MeV
energy band is leading the radio main pulse by $\sim 400 \pm 150~\mu s$. Later,
the misalignment in phase of the main radio pulse and the main pulse in shorter wavelengths
has been confirmed by several instruments. No absolute agreement exists in the value
of the radio delay measured by different instruments even for the same energy band, though they
are close to each other especially if one takes into account not only statistical errors but
also the possible systematic uncertainties. The most recent measurements
of the X,$\gamma$-rays to radio lag are: RXTE/PCA ---
$344\pm40~\mu s$ \citep{rots2004}, JEM-X/INTEGRAL --- $300\pm67~\mu s$ \citep{brandt2003},
ISGRI/ and SPI/INTEGRAL --- $285\pm12~\mu s$ and  $265\pm23~\mu s$, respectively
\citep[statistical errors only,][]{kuiper2003},
EGRET/CGRO --- $241\pm29~\mu s$ \citep{kuiper2003}.
The INTEGRAL results mentioned above are based on observations
covering only several days, thus these data can be folded with the single
ephemeris record, while different ephemerides were used for RXTE and CGRO observations.
That means that RXTE and CGRO results should be statistically more significant since
they are less affected by uncertainties in the radio ephemerides \citep[see][]{rots2004}. 
Recently, the optical-radio delay has also been confirmed, $255\pm21~\mu s$ \citep{oosterbroek2008}.

In this paper we present the timing analysis of the Crab pulsar
in the $2-300$ keV energy range
with the SPI/INTEGRAL telescope and the RXTE instruments.

% FIG 1

\begin{figure}[t]
 \vspace{-0.1cm}
 \resizebox{90mm}{!}{\hspace{-0.0cm}\includegraphics{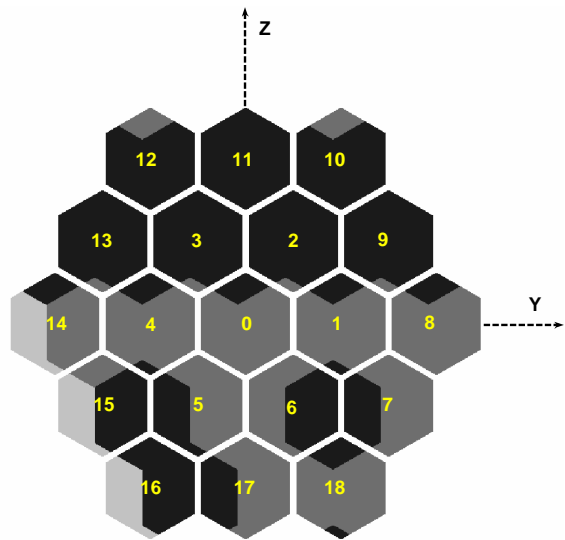}}
 \vspace{-5mm}
 \caption{The shadowgram of the detectors plane of the SPI telescope in the simple
case when only one source is located in the SPI FOV. The``black'' colour corresponds to the area
 illuminated by the source ; ``Dark grey'' to the area shadowed by the mask;
``Light grey -- to the area closed for the source by the collimator.  
}
 \label {spidets}
\end{figure}

\section{Observations and data reduction}

\subsection{INTEGRAL}

The International Gamma-Ray Astrophysics Laboratory \citep[INTEGRAL;][]{winkler03}
was launched on Proton/Block-DM on October 17, 2002
into a geosynchronous highly eccentric orbit with high perigee
\citep{eismont03}. The scientific payload
of INTEGRAL includes four telescopes: the spectrometer SPI \citep{vedrenne03},
the imager IBIS \citep{ubertini03}, the X-ray monitor JEM-X \citep{lund03}
and the optical monitor OMC \citep{mashesse03}.

\subsubsection{The SPI telescope characteristics}

In this paper, we focus on analysis of the SPI data.  For detailed description of the
instrument calibrations and performance,  see Vedrenne et al. (2003), Atti$\acute{e}$ et al. (2003) and Roques et al. (2003).
Below we give
only a brief review of the key characteristics of the instrument and software
relevant for this work.

SPI consists of 19 high
purity germanium detectors (GeD) packed into a hexagonal array (see
Figure \ref{spidets}). The combination of a good sensitivity to the
continuum and line emission in the energy band $20-8000$ keV, provided
by a large geometrical detector's plane area ($\sim500 ~cm^2$) and cryogenic
system, and a good timing resolution ($102.4 ~\mu sec$) gives us an
opportunity
to study even very fast X/$\gamma$-rays pulsars.

There are two main types of SPI events: single events (registered by only
one detector) and multiple events (scattered photon detected by two and more
detectors). Though the pulsating signal is clearly detected in both
types of events, we did not use the multiple events because of
difficulties in the extraction of the spatial information, worse
timing resolution and the lack of a tested
energy response matrix. 
It should be noted that this kind of data was successfully
used by Dean et al. (2008) for measuring Crab polarisation.

\subsubsection{The SPI clock and time connection}

SPI has its own internal clock, a 20MHz oscillator that generates time tag signal every
2048 periods, i.e. 102.4 $\mu s$ -- the time resolution of the instrument. The SPI clock is synchronised
with the on-board clock at every $125$ ms (the 8Hz on-board cycle, $OBT_{8Hz}$) by a reseting of
the counter
associated to the SPI oscillator. The SPI time tags are counted from the beginning of 
the 8Hz cycle and divide the $125$ ms interval into 1019 intervals of
102.4 micro seconds duration plus one interval with a duration of 70 micro seconds. In fact,
the base frequencies of both on-board and SPI clocks are slightly varying with time (e.g. due
to variation of the crystals temperature) that leads to variation in duration of SPI time tag
intervals. We have estimated this effect to be very 
small ($<5~\mu s$ for any statistically
significant analysis),  neglected it and used the nominal values of the
frequencies. 
The on-board time (OBT)
of the current 8Hz  cycle is the value of the datation of the beginning of the cycle
(SPI User Manual - Issue 5.2 - SEP 2002). By convention, the On Board Time (OBT) of any event is the
time of the leading edge of the interval where the event has been detected, that is :
\begin{equation}
OBT=OBT_{8Hz} + N_{tt}*D/2^{-20},
\end{equation}
where $N_{tt}$ is the
number of the SPI time tag interval ($0..1019$) and $D=102.4\times10^{-6} s$ is the length of
the interval.
The divisor $2^{-20}$ is introduced
to convert time expressed in seconds to on-board time units and reflects the conventional time
resolution of the INTEGRAL clock (the real clock accuracy is $2^{-19}$ seconds).  
For the conversion of the On Board Time to Coordinated Universal Time (UTC) and
for the barycentric correction we have used routines from the standard Off-line
Science Analysis software package version 7.0 developed at INTEGRAL Science
Data Centre \citep{courvoisier03} and the time correlation files provided with
the auxiliary data for the INTEGRAL data archive generation number 2. 
The equation (1) and the time transformation routines ensure accuracy of returning Universal
Time of the order of $100~\mu s$. For precise timing analysis, a time correction
must be added to OBT before any conversion:
\begin{equation}
\Delta T = \Delta T_{SPI} + \Delta T_{sat} + \Delta T_{rev2},
\end{equation}
where $\Delta T_{SPI}=D/2=51.2~\mu s$ --- the mean systematic shift due to the fact that the arrival
time of events is defined as the time of the leading edge of the time tag interval, while the
actual arrival times are normally distributed inside the time tag interval;\\ 
$\Delta T_{sat}=83~\mu s$ -- the delay between the on-board time and SPI time
\citep[ground calibration, ][]{alenia2002};\\
$\Delta T_{rev2}=-47~\mu s$ -- this shift is common for all time correlation files
from the archive generation 2 (will disappear in the new generation of archive).\\
Note, that the SPI instrumental delay given in the Table 4 of Walter et al. (2003)
is the sum of the first two terms ($134~\mu s$) of the expression (2).

In our analysis, to avoid an additional
discretization in time series we decided to define the time of
any registered event not as the time of the leading edge of the appropriate time tag interval,
but as a linearly randomised  time inside this time tag interval, and the final expression
for OBT is the following:
\begin{equation}
OBT=OBT_{8Hz} + \left((N_{tt}+rand[0,1])*D+\Delta T\right)/2^{-20},
\end{equation}
in this case $\Delta T_{SPI}=0$ and therefore $\Delta T = +36\times10^{-6}~s$.

%% Table 1

\begin{table}[h]
%\begin{center}
{
%\vspace{-1.0cm}
\small
\caption{List of observations used in this analysis.\label{tbl-1}}
\begin{tabular}{|l|l|c|c|}
\tableline
\tableline
{\bf } & {} & {} & {} \\
{\bf ~Rev. ~} & { Observing Period } & {Exp.$^a$} & { Target~$^b$ } \\
{~~~\bf N$^o$} & {~~~~~~~~UTC} & {ks} & {} \\
\tableline
\multicolumn{4}{|c|}{\it\bf YEAR 2003}\\
\tableline
{~\bf 0043} & {~$\bf 19.177-21.783$ \bf Feb.} & {$\bf 164$} & {\bf CRAB} \\
{~\bf 0044} & {~$\bf 22.163-24.772$ \bf Feb.} & {$\bf 177$} & {\bf CRAB} \\
{~\bf 0045} & {~$\bf 25.167-27.699$ \bf Feb.} & {$\bf 162$} & {\bf CRAB} \\
{~\bf 0102} & {~$\bf 14.623-17.218$ \bf Aug.} & {$\bf 91$} &  {\bf CRAB} \\
{~\bf 0103} & {~$\bf 17.610-17.845$ \bf Aug.} & {$\bf 22$} &  {\bf CRAB} \\
{~\bf 0124} & {~$\bf 19.430-22.031$ \bf Oct.} & {$\bf 195$} & {\bf IC443} \\
{~\bf 0125} & {~$\bf 22.422-25.023$ \bf Oct.} & {$\bf 196$} & {\bf IC443} \\
{~\bf 0126} & {~$\bf 25.414-28.015$ \bf Oct.} & {$\bf 110$} & {\bf IC443} \\
\tableline
\multicolumn{4}{|c|}{\it\bf YEAR 2004}\\
\tableline
{~\bf 0170} & {~$\bf 5.122-7.302$ \bf Mar.} & {$\bf 119$} & {\bf CRAB} \\
{~\bf 0182} & {~$\bf 9.987-12.504$ \bf Apr.} & {$\bf 197$} & {\bf IC443}  \\
{~\bf 0184} & {~$\bf 15.976-18.566$ \bf Apr.} & {$\bf 222$} & {\bf IC443} \\
{~\bf 0239} & {~$\bf 27.435-30.036$ \bf Sep.} & {$\bf 187$} & {\bf CRAB}  \\
{~\bf 0247} & {~$\bf 21.363-23.964$ \bf Oct.} & {$\bf 223$} & {\bf IC433} \\
\tableline
\multicolumn{4}{|c|}{\it\bf YEAR 2005}\\
\tableline
{~\bf 0300} & {~$\bf 28.910-31.504$ \bf Mar.} & {$\bf 188$}      & {\bf CRAB}\\
{~\bf 0352} & {~$\bf 31.463 - 2.890$ \bf Aug/Sep.} & {$\bf 200$} & \bf A0535 \\
{~\bf 0365} & {~$\bf 11.096 - 11.938$ \bf Oct.} & {$\bf 59$} & {\bf CRAB}    \\
\tableline
\multicolumn{4}{|c|}{\it\bf YEAR 2006}\\
\tableline
{~\bf 0422} & {~$\bf 28.729 - 31.198$ \bf Mar.} & {$\bf 190$} & {\bf CRAB} \\
{~\bf 0464} & {~$\bf 1.252 - 3.011$ \bf Aug.} & {$\bf 60$} & {\bf Taurus}  \\
{~\bf 0483} & {~$\bf 28.711 - 29.627$ \bf Sep.} & {$\bf 69$} & {\bf CRAB}  \\
\tableline
\multicolumn{4}{|c|}{\it\bf YEAR 2007}\\
\tableline
{~\bf 0541} & {~$\bf 19.537 - 22.128$ \bf Mar.} & {$\bf 213$} & {\bf CRAB} \\
{~\bf 0605} & {~$\bf 27.016 - 28.927$ \bf Sep.} & {$\bf 154$} & {\bf CRAB} \\
\tableline
\multicolumn{4}{|c|}{\it\bf YEAR 2008}\\
\tableline
{~\bf 0665} & {~$\bf 24.473 - 27.033$ \bf Mar.} & {$\bf 194$} & {\bf CRAB}  \\
{~\bf 0666} & {~$\bf 27.463 - 30.022$ \bf Mar.} & {$\bf 204$} & {\bf CRAB}  \\
{~\bf 0727} & {~$\bf 25.940 - 28.488$ \bf Sep.} & {$\bf 197$} & {\bf CRAB}  \\
{~\bf 0728} & {~$\bf 28.932 - 1.499$ \bf Sep./Oct.} & {$\bf 184$} & {\bf CRAB}  \\
\tableline
\multicolumn{4}{|c|}{\it\bf YEAR 2009}\\
\tableline
{~\bf 0774} & {~$\bf 13.538 -- 16.145$ \bf Feb.} & {$\bf 196$} & {\bf CRAB}  \\

% vse dannye 3.95 Ms; v tablicae -- 3.3 Ms

\tableline
\tableline

\end{tabular}

%\vspace{-0.1cm}

$^a$ -- this value represents the total exposure of selected data \\
$^b$ -- for the complete description of the observations see the ISOC site {http://www.sciops.esa.int}\\

%\vspace{-0.5cm}
}
%\end{center}
\end{table}

\subsubsection{Data selection}
In our analysis we have used publicly available data of all observations
where the source was in the FOV of the SPI telescope. The exposure of selected data
totals up approximately $4$ Ms.

SPI is a telescope with a coded mask aperture and most INTEGRAL
observations are organised as a set of snapshots of the sky around a
target (pointings or Science Windows -- continuous observations pointed on a given direction
in the sky). It means that the instrument effective area for the chosen target
is changing from pointing to pointing. This effective area is mainly determined by
geometrical area of the non-shadowed part of the detector plane (see Figure \ref{spidets}).

To extract the Crab pulsed signal we have used the epoch folding technique \citep{leahy83}.
Any set of observations can be represented as a set of the whole detector
plane count rates and in these terms, the total folded curve is the direct sum of the folded
countrates of the individual pointings.
To reach the best result we need to find an optimal series of the INTEGRAL pointings
for which the signal to noise ratio will be the highest one. In this regard,  several parameters 
characterise each pointing: the
exposure of the pointing, $T_i$; the background conditions -- the instrumental background
countrate plus the sum of the countrates from other sources in the FOV, $C_i^b$; the illumination
fraction of the detector plane,  $\alpha_i$,
corresponding to the source direction (could include not only geometrical factor, varies from
0 to $\sim0.6$); and  
the mean countrate of the pulsating part of the source emission, $C_i^p$
(for completeness, below we are introducing also the term $C_i^{dc}$ - the mean unpulsating
countrate of the pulsar, for the Crab pulsar $C_i^{dc}=0$).
Using the terminology introduced above and assuming that all variances follow
the Poisson statistic, the signal to noise ratio for the sequence of M pointings can be
expressed as follows: 

\begin{equation}
  \left(\frac{S}{N}\right)_M=\frac{\sum\limits_{i=1}^M \alpha_i C_i^{p}  T_i}{\sqrt{\sum\limits_{i=1}^M (C_i^b+\alpha_i(C_i^{p}+C_i^{dc}) ) T_i}}
\end{equation}

The optimal set of K pointings chosen from the initial set of M pointings is that for which
the value of $(S/N)_K$ calculated using the equation (4) reaches its maximum. In the case of a
large number M the exhaustive search will take infinite time (even for $M=100$ the number of combinations
exceeds $10^{30}$, in our case $M>1000$). Several simplifications have been investigated.

% FIG 2
\begin{figure}[t]
 \vspace{-0.1cm}
 \resizebox{80mm}{!}{\hspace{-0.0cm}\includegraphics{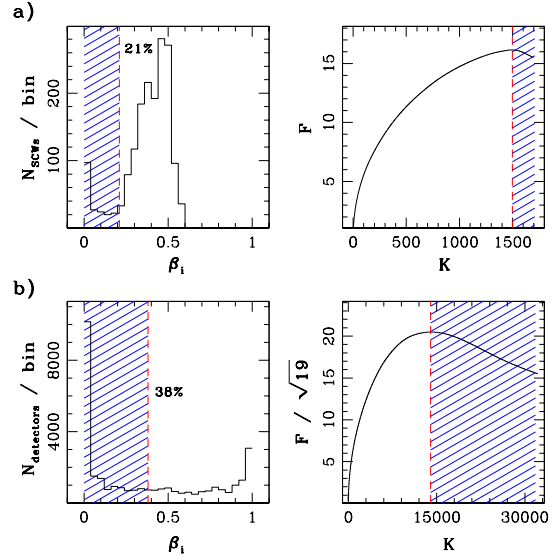}}
 \vspace{-5mm}
 \caption{The left panels show the distribution of the number of SCWs (a) or the number of individual
detectors (b) versus their illumination fraction for
our data set. The function F(K) (see the equation (6)) is plotted on the rights panels for
two cases: a) the whole detector plane; b) individual detectors. Shaded areas correspond
to the data that should be excluded from the analysis. 
The function F in the case (b) is renormalised to be in the same units as
in the case (a).
}
 \label {test}
\end{figure}

Considering that the INTEGRAL pointings (or Science Windows, SCWs) cover generally $\sim2-5$ ks time
intervals, we can treat the exposure $T_i$ of individual member of the
data series as constant, i.e. $T_i\simeq const=T$, $i\in [1,M]$. Moreover, in the SPI telescope, the
background dominates the useful signal, i.e.  $(C_i^{p}+C_i^{dc})\ll C_i^b$ is true
for any point source excluding the brightest events like Gamma-Ray Bursts or short intense bursts from
Soft Gamma-Ray Repeaters and Anomalous X-ray Pulsars. Further, the amplitude of the variation of the SPI
background does not exceed $\sim 50\%$ of its mean level (excluding periods of Solar flares).
So, we can treat $C_i^{b}\simeq const=C_{b}$
for $i\in [1,M]$.  We are assuming also that the mean value of the pulsar
``Pulsed'' component is constant in time $C_i^{p}\simeq const=C_{p}$.
Now, taking into account assumptions listed above, we can modify the equation (4) as:
\begin{equation}
  \left(\frac{S}{N}\right)_M\simeq \sqrt{\frac{T C^2_{p}}{C_b}} ~\frac{\sum\limits_{i=1}^M \alpha_i}{\sqrt{M}}
\end{equation}
In equation (5) we do not reduce the number of combinations in comparison with the equation (4)
but now it is easy to see that the procedure of searching for the optimal set is equivalent
to searching for the maximum value of the following discrete function:
\begin{equation}
F(K)=\frac{\sum\limits_{i=1}^K \beta_i}{\sqrt{K}}, K\in [1,M]
\end{equation}
where $B=[\beta_i]$ is the back ordered $A=[\alpha_i]$ set. 

%Or, the condition of the $(K+1)$'s pointing acceptance can be expressed in
%the recurrence formula:

%\begin{equation}
%F(K+1)-F(K)=\frac{\sum\limits_{i=1}^{K+1} \beta_i}{\sqrt{K+1}}-\frac{\sum\limits_{i=1}^K \beta_i}{\sqrt{K}}>0
%\end{equation}
 
The speculations presented above could be easily extended to the case
when we do not treat  the whole detector plane (hereinafter referred to as
the case I) but each detector separately (case II).
In the latter case, $\alpha_i$ is the illumination fraction of an individual detector
and varies in the range $0-1$, and M is the number of pointings times the number
of the individual detectors (19 for SPI).

We have obtained the solution for our set of INTEGRAL observations for both
cases as illustrated in Figure \ref{test}.
To reach the maximum of the signal to noise ratio in case I, we should
exclude from the analysis the pointings
with an illumination fraction below $21\%$. In case II, we should
use only those detectors that have an illumination fraction above
$38\%$.  Figure \ref{test}, shows  that using  individual detectors (case II)
we obtain a $\simeq 30\%$ improvement of the signal detection significance. In this
paper dedicated to the timing analysis, we have implemented case II. 

\subsection{Jodrell Bank Crab Pulsar Monthly Ephemeris}

For the folding of the Crab pulsar lightcurves, we have used 
the time solution derived from Jodrell Bank Crab Pulsar Monthly Ephemeris \citep{lyne93}
and the corresponding Crab Pulsar coordinates.
The database is available through the
World Wide Web (http://www.jb.man.ac.uk/pulsar/crab.html) and contains
the dispersion-corrected time of arrival of the centre of the main pulse
(in TDB time system), the frequency and its first derivative and the
range of validity. From this database we extracted the radio ephemerides
covering the periods of the INTEGRAL observations. For each radio
ephemeris record we took its two neighbours and calculated the second
derivative of the frequencies so that the phases and frequencies given at the
edge of the validity intervals 
and those deduced by extrapolation are
consistent with each other better than $0.0001$ in phase
and $10^{-7}$ in frequency. The resulting ephemerides for the INTEGRAL
observing periods are given in Table 2.

The main pulse arrival time in the monthly ephemeris is determined with
an error around
$60~\mu s$, that includes the uncertainty in the delay due to interstellar
scattering $\sim 20~\mu s$ (owing to the dispersion measure
uncertainty $\Delta DM\sim0.005~pc/cm^3$) as well as those arising
from unknown instrumental effects $\sim 40~\mu s$ (see e.g. Rots et al. 2004 and
references there). While the first part can be treated as a statistical error that
follows the Poisson statistic and decreases with the number of independent measurements,
the second part should be treated as a systematic error that is always present in the
measured values.   

% FIG 3

\begin{figure}[t]
 \vspace{-0.1cm}
 \resizebox{80mm}{!}{\hspace{-0.0cm}\includegraphics{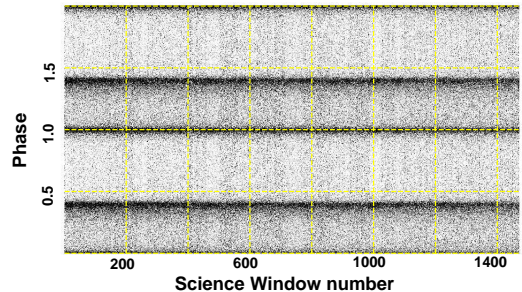}}
 \vspace{-5mm}
 \caption{Science window by Science window folded curves in the 20-100 keV energy band.
The folding procedure is based on ephemerides from the Table 2. No $\Delta \Psi$ correction is
applied (see text).
}
 \label {fold_hist}
\end{figure}

\subsection{RXTE}

The PCA instrument onboard the RXTE (Rossi X-ray
Timing Explorer) orbiting X-ray observatory consists
of five identical proportional counters with a total
area of 6500 cm$^2$, operating in the 2-60 keV energy range 
\citep{bradt93}. The accuracy of the RXTE clock in
absolute time for our observing time interval (2003-2009 yy.)
is better than $2~\mu s$ (see Rots et al. 2004 and references
therein). Because of its large area and excellent
time resolution and time accuracy, the instrument is sensitive enough
to reconstruct a significant 400 bins phase curve of the Crab pulsar
using an exposure of the order of 1 ks. We used
Crab PCA observations that coincide in time (within two weeks) with any
of our INTEGRAL observation and contain  data in the Generic event mode format
with time resolution better than $250~\mu s$. For the fine clock
correction and barycenter correction we used {\it faxbary} script
from the FTOOLS package that calls the {\it axBary} code (see e.g. the
RXTE Guest Observer Facility). For the folding procedure we use the same
routine and the same ephemerides as for SPI/INTEGRAL. 

% FIG 4

\begin{figure}[t]
 \vspace{-0.1cm}
 \resizebox{80mm}{!}{\hspace{-0.0cm}\includegraphics{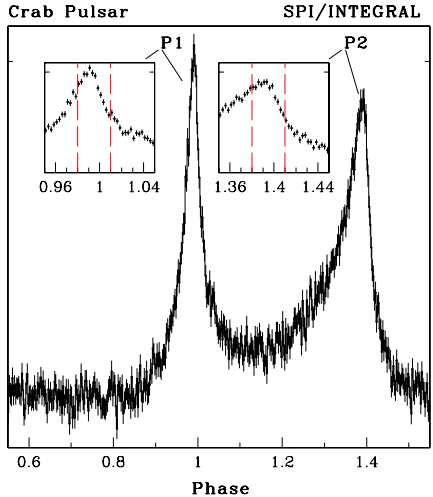}}
 \vspace{-5mm}
 \caption{Crab phase histogram in the 20-100 keV energy band in absolute
phase with the phase resolution of $0.0025$. ``1'' corresponds to the phase of the
main radio pulse. Inner panels magnify the main pulse
and the interpulse peaks. Dashed vertical lines bound the intervals used
for the fit procedure. The exposure is of the order of 100 ks.
}
 \label {crabpuls}
\end{figure}

\section{SPI/INTEGRAL analysis and results}

To perform epoch folding analysis we ascribe to each detected photon the phase $\Psi_t$
using the appropriate ephemeris from the Table 2 and the following formula:
\begin{equation}
\Psi_t=\Psi_0 + f (t-t_0) + \frac{1}{2} (t-t_0)^2 \frac{df}{dt}  + \frac{1}{6} (t-t_0)^3 \frac{d^2f}{dt^2} 
\end{equation}
where $t_0$, $f$, $\frac{df}{dt}$, $\frac{d^2f}{dt^2}$ is the radio ephemerides valid for the
moment ``t'' and $\Psi_0 \equiv 0$ (we want to work in absolute phase, i.e. the
main radio pulse is at phase $0.0$). Then we can plot the phase values producing light curves
 with the requested resolution (here we use 400 bins per
cycle). The middle of the zero bin is corresponding to the phase $0.0$.

As a first step,
to check our SPI data set for the presence of unknown ``glitches'' from the Crab
pulsar or some instrumental artifacts (e.g. inaccuracy in the on-board to Universal
time conversion procedure) we folded separately  each SCW lightcurve
in the broad $20-100$ keV energy band.
The result of the dynamical folding  is presented on Figure \ref{fold_hist},
where we see that the shape of the Crab phase histograms and absolute phases are
very stable (the two peaks of the pulse are good tracers).

In order to determine the phase of the hard X-ray main pulse and interpulse more
precisely and to study possible variations of these values in time we summed up the SCW
folded curves for each revolution. The  SPI pulse profile for the Crab pulsar
in the $20-100$ keV energy band for the exposure $\sim 100$ ks is
presented on Figure \ref{crabpuls}. To define the phase of the pulses,  we fitted 
the data for each revolution with a composite model: a Gaussian function plus a
constant background, in the phase intervals $0.98-1.01$ and $1.38-1.41$ for the main
pulse and interpulse, respectively, and adopted the fitted position of the Gaussian centroid
as the appropriate phase of the corresponding pulse. To make sure that the fit results are
model independent, we fitted the data for the main pulse with two other models:
Lorentzian --- used in Rots et al. (2004) (due to the lack of statistic for the one revolution
timescale we could not apply the complete procedure of the peak-finding described in this paper)
and Lorentzian plus constant --- used in Kuiper et al. (2003). We found that all three models
yield similar values, with dispersion not exceeding $0.0003$ period ($<10~\mu s$).
 
% FIG 5

\begin{figure}[t]
 \vspace{-0.1cm}
 \resizebox{80mm}{!}{\hspace{-0.0cm}\includegraphics{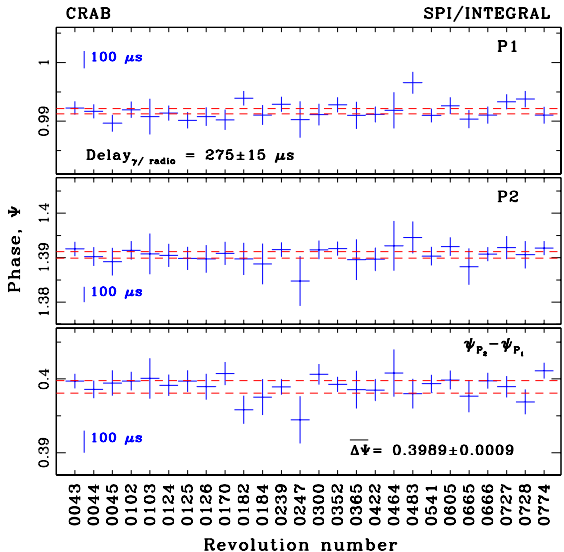}}
 \vspace{-5mm}
 \caption{The best fit values of the main pulse (P1), interpulse (P2) positions and 
distance between them in the phase plane versus the revolution number, in the $20-100$ keV energy band.
The horizontal dashed lines show the one sigma (statistical only) confidence intervals for the averaged values. 
}
 \label {timing}
\end{figure}

The absolute phases of the main pulse and interpulse (with respect to the radio main
pulse) together with the
phase difference between them, as a function of the observation number, are shown in
Figure \ref{timing}. The distributions of $\Psi_{P1}$ and $\Psi_{P2}$ values
are well consistent with normal distributions ($\Psi_{P2}-\Psi_{P1}$
is not a measured value but a combination of two independent functions). The mean value
of the hard X-ray main pulse phase is $0.99182\pm0.00046$, thus, the pulse leads the
radio main pulse by $8.18\pm0.46$ milliperiods or $275\pm15~\mu s$. Such quoted error does not
include $40~\mu s$ coming from the uncertainty of the radio ephemeris.
The average value of the two pulses separation is $0.3989\pm0.0009$ parts of the cycle.
Note, that this relative value is independent of any uncertainty in the radio timing
ephemeris. 

%%%%%%%%%%%%%%%%%%%%%%%%%%%%%%%%%%%%%%%%%%%%%%%%%%%%%%%%%%%%%%%

\section{Comparison with PCA}

\begin{figure}[t]
 \vspace{-0.1cm}
 \resizebox{80mm}{!}{\hspace{-0.0cm}\includegraphics{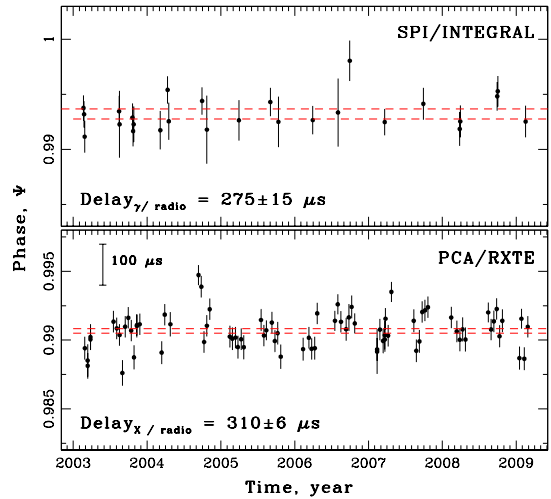}}
 \vspace{-5mm}
 \caption{The main pulse maximum  arrival phase in the $20-100$ keV
energy band with SPI/INTEGRAL (1 point by time interval order of 100 ks),
and in the $2-20$ keV energy band with PCA/RXTE
(1 point ---  a few kiloseconds).
The same radio ephemerides and fit procedure have been used. The horizontal
dashed lines show the one sigma (statistical only) confidence intervals for the averaged values.
}
 \label {pca_spi}
\end{figure}

For independent check of the hard X-ray results, we carried out the analogous
analysis in the 2-20 keV energy band. We used the data of Crab observations
with the PCA monitor quasisimultaneous
with INTEGRAL. We used the
same radio ephemeris and the same pulse definition procedure. The phase positions of the X-ray main
pulse relative to the radio one  for 79 PCA/RXTE observations
are shown in the bottom panel of Figure \ref{pca_spi}, showing that also the main X-ray
pulse leads the radio one. To determine the mean value of the time lag we approximated
the PCA data with a constant and got $0.00933\pm0.00016$ in phase units or
$310\pm6 \mu s$ in time. Again we are quoting only statistical errors but we keep in mind the
systematic
$\sim 40 \mu s$ error that comes from the radio ephemeris uncertainties.
Based on two measured time lags ``X-ray/radio'' and ``hard X-ray/radio'' we can conclude
that the main X-ray pulse is leading the hard X-ray main pulse by $35 \pm 16 \mu s$.
This value differs only marginally from zero (we provided the one sigma error)
even taking into account that in this case the radio error does not play a role
since we used the same radio ephemerides for both measurements. Another quantity that is
independent of any uncertainty in the radio timing ephemerides is the phase difference
between the main pulse and interpulse. From our set of PCA/RXTE observations, we
got a value of $0.40016\pm0.00028$ that is in a good agreement with the result obtained
in hard X-rays ($0.3989\pm0.0009$, SPI/INTEGRAL). 

% FIG 6

\section{Radio delay evolution with energy}

\begin{figure}[t]
 \vspace{-0.1cm}
 \resizebox{80mm}{!}{\hspace{-0.0cm}\includegraphics{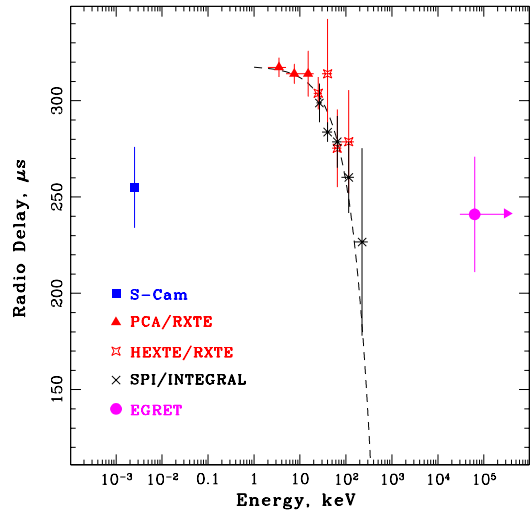}}
 \vspace{-5mm}
 \caption{The radio delay in function of energy. The optical (S-Cam) and $\gamma-$ray
(EGRET) points are from Oosterbroek et al. 2008 and Kuiper et al. 2003, respectively.
The data of the RXTE instruments and SPI are from this work.
}
 \label {endep}
\end{figure}

We have measured  accurately the delays of the main pulse arrival time in wide $2-20$ keV
(soft X-rays) and
$20-100$ keV (hard X-rays) energy bands with respect to the radio main pulse arrival time. 
The $310\pm6~\mu s$ soft X-ray/radio delay derived in this paper based on
the PCA data 
(Rots et al. 2004 provides even higher value $344\pm40~\mu s$) is marginally higher than the
radio delay with respect to $275\pm15~\mu s$ hard X-ray/radio delay measured with SPI. Both
values are also slightly higher of the radio/optical delay of $255 \pm 21~\mu s$ derived
from S-Cam optical observations
(Oosterbroek et al. 2008) and the $241\pm29~\mu s$ radio/gamma one ($>30$ MeV, 
EGRET, Kuiper et al. 2003). To check whether this differences are real or not, we have
made an additional analysis and investigated the behaviour of the radio delay with energy. 
The radio delay evolution with energy have been observed in optical wavelength
(Oosterbroek et al. 2008),
that gave us an extra motivation.
We have built folded curves in narrower
energy bands. We split the $2-20$ keV PCA energy band on three parts, while
we used five energy channels to cover the 20-300 keV  energy band for SPI data.
We also added the High-Energy X-Ray Timing Experiment (HEXTE/RXTE) data in four energy bands
covering the $20-250$ keV energy range, allowing a direct crosschecking
with SPI results. The position of the main peak has been determined as previously,
from a fit with a composite model: a Gaussian function plus a
constant background, in the phase interval  $0.98-1.01$.
Figure \ref{endep} presents the evolution of the radio delay versus energy for
the RXTE and SPI data, together with the optical and $\gamma-$rays points from
Oosterbroek et al. (2008) and Kuiper et al. (2003). The decreasing trend of the radio delay
with energy in the ($2-300$ keV) energy domain supports the reality of
the delay measured between the soft X- and hard X-ray main peaks, even though the
individual error bars are large. When modelling this decrease by a simple linear law 
(dashed line in Figure \ref{endep}  )
we find that the radio delay decreases  with a rate of $\sim 0.6 \pm 0.2 ~\mu s/keV$.
The obtained Chi2 of  7.1 for 10 dof compared to the value of 42.4 for 11 dof
for a constant model corresponds to a probability of 3.5$\times 10^{-5}$ that 
its improvement is by chance.

% FIG 7

\section{Discussion and conclusions}

We have investigated the pulse profile of the Crab pulsar between 2 and 300 keV with
PCA/RXTE instrument and SPI/INTEGRAL telescope.
We found the strong indication that in this energy range the radio delay is
significantly decreasing with energy.
The simplest explanation of such a behaviour is that the $X-ray/\gamma_{soft}$-ray emission
originates in a region extended along the open magnetic field lines, with 
softer photons  originating at higher altitudes while the time offsets represent simply the
path-length differences. In this case, the time delay determines the characteristic azimuthal size
of the emitting area as $54\pm18$ km. 

It is clear that the radio delay can not decrease
linearly  through the full gamma rays range,
since the EGRET point imposes a positive value.
Moreover, contrary to we observe above 2 keV,
the radio delay increases between optical and X-rays. It is interesting to note that
the X-ray delay is
consistent with the rate of $-5.9\pm1.9 ~\mu s/1000~\AA$ derived from
the $3920-8230~\AA$ optical waveband by Oosterbroek et al. (2008). 

Even thought the dependence of the main pulse phase position on energy is
complex, we can explain it with a rather simple scheme, where the pulsed emission consists of 
the superposition of two independent components
having different phase distributions and energy spectra. Indeed, such an analytical model
has been introduced by Massaro et al. (2000) to interprete the BeppoSAX data in
the 0.1-300 keV energy band \citep[see also improvements][]{massaro2006,campana2009}.
In the optical up to $\gamma-$ray domain, this model includes: an ``optical'', $C_o$
and an ``X-ray'', $C_x$.  
The fractional part of the $C_x$ component increases in the main pulse with increasing energy
from 1 keV up to a 1 MeV, then decreases  up to $\sim 10$ MeV energy but is negligible below 1 keV 
and above 10 MeV. $C_x/C_o$ ratio behaviour following
the same law as  the Bridge/P1 ratio presented in Kuiper et al. (2001).
Thereby, in the X-ray band,
the $C_x$ component shifts the maximum of the pulse I $(C_x+C_o)$ emission toward
the radio maximum on the phase plane. It explains that the radio delay decreases with energy in X-rays
while it keeps
identical values in the optical and $\gamma-$ray wavebands where the $C_x$ component is
negligible. Considering that the values of the radio delay
are nearly the same in the optical wavelengths and $\gamma-$rays, where $C_o$ largely dominates over $C_x$, 
we can suggest that the $C_o$ component corresponds to a single emission mechanism and emission location
from optical to $\gamma-$rays. In this case, the $\sim 250~\mu s$ radio delay indicates that
the radio emission is produced in a region located closer from the Neutron Star by $\sim 75$ km than the $C_o$
production site.
 On the other hand, the $C_x$ component could be unrelated to the $C_o$ one, with a different
origin and/or source location. In this case, the source
behaviour in X-rays would  result from a superposition of (at least) two independent components and
the  measured values of the radio delay in this energy band would not have any direct physical
explanation. 
%Moreover, since the main pulse shape evolving with energy, these precise values are
%model dependent. 

\acknowledgments

We are grateful to Prof. Hermsen, Dr. Kuiper, Dr. Rots and Dr. Revnivtsev for the very helpful
technical notes and discussion. We wish to thank to the INTEGRAL helpdesk team (especially
Dr. Turler, ISDC) and Dr. Southworth (ESOC) for productive discussion of the INTEGRAL timing
facilities and prompt correction of newly discovered problems. We acknowledge HEASARC at NASA/Goddard
Space Flight Center for maintaining its online archive service which provided the RXTE
data used in this paper. This research has been supported by CNES.

\onecolumn

%% Table 2

\begin{table}[h]
\small
\caption{The Crab pulsar ephemerides for the INTEGRAL observations listed in Table \ref{tbl-1}.
We used the Crab Pulsar position given in Jodrell Bank:  $RA=83.633217^{\circ}$ and $DEC=22.014464^{\circ}$ 
in the J2000 epoch\label{tbl-2}}
\begin{tabular}{|l|c|c|c|l|c|c|c|}
\tableline\tableline
{}               &  $$    & $$ & $$ & {} &  $$    & $$ & $$\\
{\bf ~Rev. ~} & $T_{valid}$ (MJD), & $~f~~~~$, {~~~~Hz~~~~} & $\Delta t_0, \mu sec$ & {\bf ~Rev. ~} &  $T_{valid}$ (MJD), & $~f~~~~$, {~~~~Hz~~~~} & $\Delta t_0, \mu sec$\\
{{~~~\bf N$^o$}} &  {$t^{Int}_{0}$ (MJD),} & $~df/dt~~$,  $10^{-10}$ sec$^{-2}$    & $\Delta\Psi$ & {{~~~\bf N$^o$}} &  {$t^{Int}_{0}$ (MJD),} & $~df/dt~~$,  $10^{-10}$ sec$^{-2}$    & $\Delta\Psi$\\
{}               &  $t^{MP}_0$ (sec)    & $d^2f/dt^2$, $10^{-21}$ sec$^{-3}$ & $$ & {}             &  $t^{MP}_0$ (sec)  & $d^2f/dt^2$, $10^{-21}$ sec$^{-3}$ & $$\\
{}               &  $$    & $$ & $$ & {} &  $$    & $$ & $$\\
\tableline
{~\bf }     &  $52671-52699$  & $29.8092705147$ & $18$   & {~\bf }     &  $53431-53461$      & $29.7847841837$ & $-19$\\
{~\bf 0043} &  $52685$        & $-3.7366060$    & $0.0005$ & {~\bf 0300} &  $53444$            & $-3.7315793$    & $-0.0006$\\
{~\bf }     &  $0.076659$     & $9.0$           & $$        & {~\bf }     &  $0.033023$         & $6.0$           & $$\\
\tableline
{~\bf }     &  $52671-52699$  & $29.8092705147$ & $-1$   & {~\bf }     &  $53584-53615$      & $29.7798524524$ & $36$\\
{~\bf 0044} &  $52685$        & $-3.7366060$    & $-0.0000$ & {~\bf 0352} &  $53597$            & $-3.7299236$    & $0.0011$\\
{~\bf }     &  $0.076659$     & $9.0$           & $$        & {~\bf }     &  $0.029626$         & $9.0$           & $$\\
\tableline
{~\bf }     &  $52671-52699$  & $29.8092705147$ & $-69$  & {~\bf }     &  $53644-53675$      & $29.7778867428$ & $-24$\\
{~\bf 0045} &  $52685$        & $-3.7366060$    & $-0.0021$ & {~\bf 0365} &  $53658$            & $-3.7294045$    & $-0.0007$\\
{~\bf }     &  $0.076659$     & $9.0$           & $$        & {~\bf }     &  $0.022656$         & $4.0$           & $$\\
\tableline
{~\bf }     &  $52852-52883$  & $29.8034282349$ & $8$    & {~\bf }     &  $53796-53826$      & $29.7730221322$ & $-19$\\
{~\bf 0102} &  $52866$        & $-3.7350193$    & $0.0002$  & {~\bf 0422} &  $53809$            & $-3.7278153$    & $-0.0006$\\
{~\bf }     &  $0.018825$     & $-1.0$          & $$        & {~\bf }     &  $0.005391$         & $4.0$           & $$\\
\tableline
{~\bf }     &  $52852-52883$  & $29.8034282349$ & $31.0$    & {~\bf }     &  $53917-53948$      & $29.7690932051$ & $4$\\
{~\bf 0103} &  $52866$        & $-3.7350193$    & $-0.0009$  & {~\bf 0464} &  $53931$            & $-3.7267461$    & $0.0001$\\
{~\bf }     &  $0.018825$     & $-1.0$          & $$        & {~\bf }     &  $0.020193$         & $15.0$          & $$\\
\tableline
{~\bf }     &  $52913-52944$  & $29.8014598690$ & $-11$    & {~\bf }     &  $53971-54009$      & $29.7670971393$ & $162$\\
{~\bf 0124} &  $52927$        & $-3.7344076$    & $-0.0003$  & {~\bf 0483} &  $53993$            & $-3.7264807$    & $0.0048$\\
{~\bf }     &  $0.019997$     & $3.0$           & $$        & {~\bf }     &  $0.011812$         & $18.0$          & $$\\
\tableline
{~\bf }     &  $52913-52944$  & $29.8014598690$ & $-52$    & {~\bf }     &  $54160-54191$      & $29.7612711958$ & $-25$\\
{~\bf 0125} &  $52927$        & $-3.7344076$    & $-0.0016$  & {~\bf 0541} &  $54174$            & $-3.7245631$    & $-0.0007$\\
{~\bf }     &  $0.019997$     & $3.0$           & $$        & {~\bf }     &  $0.020652$         & $5.0$           & $$\\
\tableline
{~\bf }     &  $52913-52944$  & $29.8014598690$ & $-31$     & {~\bf }     &  $54344-54374$      & $29.7553516229$ & $31$\\
{~\bf 0126} &  $52927$        & $-3.7344076$    & $-0.0009$  & {~\bf 0605} &  $54358$            & $-3.7225368$    & $0.0009$\\
{~\bf }     &  $0.019997$     & $3.0$           & $$        & {~\bf }     &  $0.010645$         & $13.0$          & $$\\
\tableline
{~\bf }     &  $53068-53074$  & $29.7969173441$ & $-50.0$   & {~\bf }     &  $54526-54557$      & $29.7494993041$ & $-46.0$\\
{~\bf 0170} &  $53071$        & $-3.7535080$    & $-0.0015$ & {~\bf 0665} &  $54540$            & $-3.7206976$    & $-0.0014$\\
{~\bf }     &  $0.010486$     & $350.0$         & $$        & {~\bf }     &  $0.008472$         & $13.0$          & $$\\
\tableline
{~\bf }     &  $53095-53105$  & $29.7958809510$ & $73$   & {~\bf }     &  $54526-54557$      & $29.7494993041$ & $-23$\\
{~\bf 0182} &  $53100$        & $-3.7414057$    & $0.0022$ & {~\bf 0666} &  $54540$            & $-3.7206976$    & $-0.0007$\\
{~\bf }     &  $0.008622$     & $200$           & $$        & {~\bf }     &  $0.008472$         & $13.0$          & $$\\
\tableline
{~\bf }     &  $53105-53115$  & $29.7955577865$ & $-22$   & {~\bf }     &  $54710-54741$      & $29.7435856030$ & $54$\\
{~\bf 0184} &  $53110$        & $-3.7392771$    & $-0.0007$ & {~\bf 0727} &  $54724$            & $-3.7189687$    & $0.0016$\\
{~\bf }     &  $0.024904$     & $-100$          & $$        & {~\bf }     &  $0.013181$         & $17.0$          & $$\\
\tableline
{~\bf }     &  $53255-53279$  & $29.7906210263$ & $40$   & {~\bf }     &  $54710-54741$      & $29.7435856030$ & $69$\\
{~\bf 0239} &  $53263$        & $-3.7333983$    & $0.0012$  & {~\bf 0728} &  $54724$            & $-3.7189687$    & $0.0021$\\
{~\bf }     &  $0.021234$     & $7.0$           & $$        & {~\bf }     &  $0.013181$         & $17.0$          & $$\\
\tableline
{~\bf }     &  $53279-53311$  & $29.7896533932$ & $-48.0$   & {~\bf }     &  $54863-54892$      & $29.7386704404$ & $-23$\\
{~\bf 0247} &  $53293$        & $-3.7329692$    & $-0.0014$ & {~\bf 0774} &  $54877$            & $-371735.72$    & $-0.0007$\\
{~\bf }     &  $0.101225$     & $9.9$           & $$        & {~\bf }     &  $0.000804$         & $14.9$          & $$\\
\tableline

\tableline
\tableline

\end{tabular}

\end{table}
%}

%%%%%%%%%%%%%%%%%%%%%%%%%%%%%%%%%%%%%%%%%%%%%%%%%%%%%%%%%%%%%%%

\end{document}